\documentclass{article}

\usepackage[margin=1in]{geometry}
\usepackage{setspace}
\usepackage[bottom]{footmisc}
\usepackage{amsmath,amsfonts,amssymb}
\usepackage{graphicx}
\usepackage{float}
\usepackage{url}
\usepackage{xcolor}
\usepackage{bm}
\usepackage{lscape}
\usepackage{siunitx}
\usepackage{colortbl}
\usepackage{threeparttable}
\definecolor{lightgray}{gray}{0.9}
\usepackage[T1]{fontenc}
\usepackage[utf8]{inputenc}
\usepackage{xcolor}
\definecolor{RoyalBlue}{RGB}{65,65,225}

\makeatletter
\renewcommand{\maketitle}{%
  \begin{center}
    {\LARGE\bfseries \@title \par}
    \vskip 1.5em

    {\large
    Niloofar Ramezani\textsuperscript{1} \quad and \quad Jeffrey R. Wilson\textsuperscript{2}
    \par}
    \vskip 0.8em

    {\normalsize
    \textsuperscript{1}Department of Biostatistics, Virginia Commonwealth University\\
    \textsuperscript{2}Department of Economics, Arizona State University
    \par}
    \vskip 0.8em

    {\normalsize
    \textsuperscript{1}{\color{RoyalBlue}\texttt{ramezanin2@vcu.edu}} \quad\quad
    \textsuperscript{2}{\color{RoyalBlue}\texttt{jeffrey.wilson@asu.edu}}
    }
    \vskip 2em
  \end{center}
}
\makeatother

\title{Semi-partitioned Generalized Method of Moments for Longitudinal Data with Lagged and Feedback Covariates}

\date{}

\begin{document}
\maketitle

\begin{abstract}
We propose a semi-partitioned Generalized Method of Moments (GMM) framework for analyzing longitudinal data with time-dependent covariates, within a marginal modeling paradigm. This approach addresses limitations of both aggregated and fully partitioned GMM models. Aggregated methods obscure temporal dynamics by assuming constant effects, while fully partitioned approaches offer temporal specificity at the cost of increased model complexity and instability—particularly with moderate sample sizes or deep lag structures. Our method distinguishes immediate from lagged effects by estimating contemporaneous coefficients separately and grouping lagged moment conditions into structured sets, while retaining flexibility in the lag‑specific effects. This yields a model that is both statistically efficient and interpretable, capturing essential temporal variation while mitigating variance inflation and convergence challenges associated with full partitioning. The framework accommodates feedback, supports both continuous and binary outcomes, and utilizes the Broyden–Fletcher–Goldfarb–Shanno (BFGS) algorithm for reliable optimization. Through simulations, we demonstrate that the semi-partitioned GMM achieves coverage and competitive efficiency relative to fully partitioned models when the grouped‑lag structure approximates the underlying lag pattern. Applications to clinical datasets on knee osteoarthritis and adolescent obesity confirm that the method recovers consistent, interpretable effects and avoids instability associated with finely grained partitioning.
\end{abstract}

\vspace{0.2in}
\section{Introduction}
Longitudinal studies, in which individuals are measured repeatedly over time, are essential for understanding dynamic processes in medical, behavioral, and epidemiological research. In the presence of multiple measurements for subjects of a study and the interest of patterns of change over time, longitudinal data are formed with the main characteristics of dependence among repeated measurements per subject \cite{zeger1986longitudinal}. However, the analytical complexity of such data arises from the correlation among repeated measures within subjects and from the time-varying nature of both covariates and outcomes. Traditional cross-sectional models fail to account for these dependencies, potentially leading to biased inference and inefficient estimation \cite{bena2008survival}. 

Marginal models—such as those based on Generalized Estimating Equations (GEE) or the Generalized Method of Moments (GMM)—have become foundational tools for estimating population-averaged effects in longitudinal settings. These approaches are appropriate when population-average inferences are the primary focus \cite{diggle2024longitudinal, hansen2010generalized} or when expected response values based on current covariates are the key quantities of interest \cite{sullivan1994cautionary}. These models focus on modeling the marginal mean of the response without requiring explicit modeling of within-subject variability, making them particularly attractive when the research objective centers on overall population trends rather than individual trajectories \cite{lee2004conditional, zorn2001generalized}. In this work, we operate explicitly within a marginal mean modeling framework: the expected outcome at time $t$ is modeled as a function of current and lagged covariates, without conditioning on past outcomes or subject-specific random effects. This contrasts with conditional mixed or autoregressive models, which specify the evolution of the outcome given its history.

A persistent and understudied challenge in marginal modeling is the appropriate handling of time-dependent covariates. These covariates can exhibit feedback from the outcome, vary in association strength over time, or exert both immediate and delayed effects. If improperly handled, time-dependent covariates can invalidate standard estimating equations and bias marginal effect estimates. Lai and Small \cite{lai2007marginal} provided a seminal classification of such covariates into three types (I, II, and III) based on their relationships to past and future responses, laying a foundation for more principled estimation using moment conditions. While certain conditional time‑series models (e.g., mixed‑data sampling (MIDAS)) also aggregate lagged covariates, those approaches target conditional expectations and forecasting, whereas our focus remains on marginal mean structures for repeated‑measures data.

Building on this classification, two major modeling strategies have emerged within the GMM framework. Aggregated GMM methods combine all valid moment conditions to estimate a single coefficient per covariate. While computationally efficient, this approach could mask meaningful temporal variation. At the other extreme, fully partitioned GMM models treat each lag of the covariate separately, offering the most flexible temporal resolution. However, they introduce several additional parameters, especially with longer follow-up durations, and often suffer from inflated standard errors, singularity problems, or convergence failure—particularly when sample sizes are not large enough.

In response to these limitations, we introduce a semi-partitioned GMM model that strategically balances flexibility and efficiency. Our approach separates the estimation of contemporaneous (lag-0) effects from lagged effects and allows lagged moment conditions to be grouped into interpretable blocks—such as short-term versus long-term lags or as a single average effect. This structure retains the richness of the moment condition framework while reducing the estimation burden and avoiding unnecessary model complexity. 

Unlike fully partitioned approaches \cite{irimata2019partitioned}, which estimate one coefficient per lag, our model mitigates variance inflation and can improve computational stability in settings where full partitioning becomes unstable by limiting the number of parameters to be estimated. At the same time, it extends beyond aggregated methods \cite{lai2007marginal, lalonde2014gmm} by explicitly modeling the structure of temporal dependencies rather than collapsing them. Furthermore, the semi-partitioned model accommodates feedback processes (Type III covariates) and supports both continuous and binary outcomes. Estimation is performed using the Broyden–Fletcher–Goldfarb–Shanno (BFGS) algorithm, a quasi-Newton method that offers improved numerical performance compared to derivative-free approaches previously used in GMM applications.

The semi-partitioned specification is not intended as a modeling endpoint but as a pragmatic compromise between flexibility and stability. Fully partitioned GMM remains valuable as an exploratory diagnostic tool for identifying lag-specific patterns, which can subsequently inform more parsimonious semi-partitioned structures. 

The remainder of this paper is organized as follows. Section 2 reviews the GMM framework and the existing aggregated and fully partitioned approaches. Section 3 presents the semi-partitioned GMM model in detail, including its mathematical formulation and estimation strategy. Section 4 provides simulation studies that evaluate model performance under varying lag structures and feedback dynamics. Section 5 presents two real-world applications: a continuous outcome from a study on knee osteoarthritis, and a binary outcome from a longitudinal study on adolescent obesity. Section 6 discusses strengths, limitations, and avenues for future research, and Section 7 concludes. 

\section{Generalized Method of Moments}

The GMM, originally introduced by Hansen \cite{hansen1982} in the context of econometrics, provides a powerful and flexible estimation framework that is particularly useful in longitudinal data settings \cite{hall2005generalized}. GMM relies on moment conditions—expected values of functions of the data and parameters—that hold under the true data-generating process, without requiring full specification of the underlying distribution. This semi-parametric approach is advantageous in longitudinal analyses, where distributional assumptions are often difficult to verify, and time-dependent covariates complicate likelihood-based methods. 
Let $\bm{\beta}_{0}$ denote the true parameter vector ans $\mathbf{x}_{it}$ represent observed covariates for subject $i$ at time $t$, with $i=1,\dots,n$ and $t=1,\dots,T$. In the marginal modeling context considered here, these moment conditions characterize population-averaged relationships between the outcome and covariates, distinguishing our setting from conditional time-series models that specify the full dynamic evolution of the outcome and distributional assumptions.

The GMM estimation procedure begins with a set of population moment conditions: 
\[
E[f(\mathbf{x}_{it}, \bm{\beta}_{0})] = 0,
\]
where $f(.)$ is a vector of functions corresponding to the moment conditions. The sample analog is formed by averaging over subjects: 
\[
\frac{1}{n} \sum_{i=1}^{n} f(\mathbf{x}_{it},\bm{\beta}).
\]
The GMM estimator minimizes a quadratic form of the sample moment vector:
\[
Q(\bm{\beta}) = \left( \frac{1}{n} \sum_{i=1}^{n} f(\mathbf{x}_{it}, \bm{\beta}) \right)^{\top} 
\bm{W} 
\left( \frac{1}{n} \sum_{i=1}^{n} f(\mathbf{x}_{it}, \bm{\beta}) \right)
\]
where $\bm{W}$ is a positive semi-definite weighting matrix. In the optimal two-step GMM, $\bm{W}$ is taken as a consistent estimate of 
$\bm{S}^{-1}$, where $\bm{S}$ denotes the long-run covariance matrix of the 
sample moment conditions \cite{hansen1982}. The minimizer of $Q(\bm{\beta})$, denoted $\hat{\bm{\beta}}$ is the GMM estimate of $\bm{\beta}_{0}$: 
\begin{equation}
    \hat{\bm{\beta}} = {\arg\min}\; Q(\bm{\beta}),
\end{equation}
where $\arg\min$ denotes the value of $\bm{\beta}$ that minimizes the function. In linear models, $\hat{\bm{\beta}}$ may be available in closed form; in nonlinear models, numerical optimization is required.

\subsection{Aggregated GMM}
Lai and Small \cite{lai2007marginal} extended the GMM framework to accommodate time-dependent covariates in longitudinal marginal regression models. Central to their approach is a classification of covariates into three types—Type I, Type II, and Type III—based on the direction and nature of association between the covariates and the outcome process over time. 

Type I covariates are those that are uncorrelated with both past and future values of the response. These covariates are typically exogenous and unaffected by outcome history, making them the most straightforward to handle in marginal models. Examples include age, calendar time, or randomized treatment assignment in crossover trials. Moment conditions involving Type I covariates are valid across all time points, and thus provide the richest set of estimating equations.

Type II covariates may be correlated with past outcomes but are assumed to be conditionally independent of future responses. This structure frequently arises in autoregressive settings or in longitudinal designs where covariates reflect evolving subject-level characteristics influenced by previous health states but not by future outcomes. Moment conditions for Type II covariates are valid for current and future responses but not for past ones, yielding a reduced—but still informative—set of estimating equations.

Type III covariates introduce feedback from the response to the covariate process itself. In these cases, the covariate at time $t$ may be a function of prior outcomes, violating exogeneity assumptions. This dynamic is common in studies of behavior or medication adherence, where responses can influence future exposure. Only contemporaneous moment conditions are valid for Type III covariates, significantly limiting the number of usable estimating equations.

Aggregated GMM approaches combine all valid moment conditions available for a given covariate into a single estimating equation. This results in one coefficient per covariate, effectively assuming that its effect is constant across time and across all valid lags. This pooling strategy simplifies estimation and often improves efficiency. However, it may obscure important lag-specific dynamics and limit interpretability. When covariate effects are heterogeneous or exhibit delay or feedback, aggregation may be less desirable.

\subsection{Fully Partitioned GMM}

To address the limitations of aggregation, Irimata et al. \cite{irimata2019partitioned} proposed a fully partitioned GMM framework that estimates a separate regression coefficient for each lag of each time-dependent covariate. Rather than assuming a constant effect over time, this approach explicitly models the lag structure by associating each observed covariate value at a given lag with a distinct regression parameter. This enables a detailed and flexible representation of how covariate effects evolve across time, providing insight into both immediate and delayed associations with the response.

The fully partitioned GMM model operates by restructuring the covariate matrix into a lower-triangular form, where each row corresponds to a response at a given time point and each column to a lagged value of the covariate. For a covariate measured at times $s=1,\dots,t$, this design matrix allows estimation of time-specific effects, denoted $\beta^{[0]}, \beta^{[1]}, \beta^{[2]}, \dots, \beta^{[T-1]}$ corresponding to contemporaneous and lagged values of the covariate. In this framework, the systematic component of the model becomes a weighted sum of these lag-specific contributions, and the moment conditions are partitioned accordingly to reflect the assumed temporal structure.
By not pooling across time, the fully partitioned GMM can uncover lag-specific variation in covariate effects that would be obscured in an aggregated model. This is particularly useful in clinical or behavioral studies where intervention effects, exposure responses, or adherence behaviors may vary over time or depend on recent history.

However, the increased flexibility comes with substantial computational and statistical cost. Partitioning increases the number of parameters in proportion to the number of time points and covariates, leading to a much higher-dimensional estimation problem. In practice, this can result in numerical instability, non-convergence, or unreliable estimates, especially when the number of subjects is not substantially larger than the number of lag-specific parameters. These problems are exacerbated in models with binary outcomes, where sparse event counts further reduce the effective sample size for each lag.
As a result, while the fully partitioned GMM offers advantages for exploring temporal heterogeneity, it is best suited to studies with relatively few time points or very large sample sizes. In typical applied settings with moderate follow-up and limited power, the fully partitioned approach may be infeasible or lead to overfitting, motivating the development of more parsimonious alternatives such as the semi-partitioned model. In practice, fully partitioned GMM can also serve as an exploratory diagnostic tool for identifying lag-specific patterns that may subsequently inform parsimonious semi-partitioned specifications.
The semi-partitioned approach introduced next is designed to retain the interpretability of lag-specific modeling while avoiding the instability and dimensionality challenges inherent in the fully partitioned specification.

\section{Semi-Partitioned GMM}

We propose a semi-partitioned GMM framework that blends the strengths of both aggregated and fully partitioned GMM models. While aggregated GMM assumes time-constant covariate effects and fully partitioned GMM models each lag separately, the semi-partitioned approach strikes a balance by estimating the immediate (contemporaneous) effect of a covariate separately from its lagged effects, which are grouped into one or more predefined structures. These predefined structures allow a controlled representation of lagged effects—whether as a combined average or categorized by relative temporal proximity (e.g., near vs. distant lags). This method is especially suited to longitudinal datasets with a moderate number of follow-up periods, where over-parameterization may hinder the convergence and stability of fully partitioned models. Furthermore, the method accommodates feedback dynamics by allowing outcomes to inform future covariate structures. Grouping lagged covariates provides a parsimonious way to summarize temporal structure, reducing dimensionality without requiring the underlying lag‑specific effects to be equal.

We adopt a grouped‑lag organizational structure to reduce the dimensionality of the moment conditions. This structure incorporates information from multiple lags while retaining flexibility in the lag‑specific effects. Efficiency gains are most pronounced when the grouping aligns with the underlying temporal pattern, and our simulations indicate that performance remains stable even under moderate misalignment.

\vspace{1em}
{\bf Fully Partitioned Model.}
Let $Y_{it}$ be the outcome at time $t$ for subject $i$, where $i=1,\dots,n$ and $t=1,\dots,T$, and let $x_{ij(s)}$ denote the value of covariate $j$ at time $s$, with $s = t, t-1, \dots, 1$. For each covariate $j$, the fully partitioned model incorporates both immediate and lagged effects as:
\begin{equation}
g(\mu_{it}) 
= \beta_0 
+ \sum_{j=1}^{J} \left( 
\beta_j^{|0|} x_{ij(t)} 
+ \sum_{k=1}^{t-1} \beta_j^{|k|} x_{ij(t-k)} 
\right),
\end{equation}
where $\mu_{it} = E[Y_{it} \mid x_{ij(s)}]$, $g(\cdot)$ is an appropriate link function, $\beta_j^{|0|}$ is the immediate effect, and $\beta_j^{|k|}$ denotes the lag-$k$ effect for covariate $j$.

\vspace{1em}
{\bf General Semi-Partitioned Model.}
To improve parsimony and stabilize estimation, lagged effects may be grouped so that all lags within a block share a common coefficient. Let $\{G_{j1},\dots,G_{jm_j}\}$ denote a partition of the lags for covariate $j$, where $m_{j}$ is the number of elements in the partition for the $j$th covariate. A general grouped-lag specification can be written as:
\begin{equation}
g(\mu_{it})
= \beta_0
+ \sum_{j=1}^{J} \left(
\beta_j^{|0|} x_{ij(t)}
+ \sum_{g=1}^{m_j} \beta_{jg} \sum_{k \in G_{jg}} x_{ij(t-k)}
\right),
\end{equation}
where $G_{jg}$ is the $g$th group of lag indices and $\beta_{jg}$ is the common coefficient for all lags $k \in G_{jg}$.

\vspace{1em}
{\bf Special Cases of Semi-Partitioned Model.} More flexible semi-partitioned structures can also be used. For instance, if “near” lags 
(e.g., $k = 1,\dots,t-r$) are grouped separately from “far” lags 
(e.g., $k = t-r+1,\dots,t-1$), the model becomes:
\begin{equation*}
g(\mu_{it}) 
= \beta_0 
+ \sum_{j=1}^{J} \left(
\beta_j^{|0|} x_{ij(t)}
+ \beta_{j,\text{near}} \sum_{k=1}^{t-r} x_{ij(t-k)}
+ \beta_{j,\text{far}} \sum_{k=t-r+1}^{t-1} x_{ij(t-k)}
\right).
\end{equation*}

As a specific case, if the first lag is believed to have a distinct effect while all remaining lags share a common effect, the first lag is modeled separately and the remaining lags are grouped together, yielding the model:
\begin{equation}
\label{eq:groupedlag}
g(\mu_{it}) 
= \beta_0 
+ \sum_{j=1}^{J} \left(
\beta_j^{|0|} x_{ij(t)}
+ \beta_j^{|1|} x_{ij(t-1)}
+ \beta_{j,\text{post-1}} \sum_{k=2}^{t-1} x_{ij(t-k)}
\right).
\end{equation}

A final special case, used in the simulation study, assumes that all lagged effects for 
covariate $j$ are summarized by a single block:
\begin{equation}
\label{eq:groupedlag2}
g(\mu_{it}) 
= \beta_0 
+ \sum_{j=1}^{J} \left(
\beta_j^{|0|} x_{ij(t)}
+ \beta_{j,\text{lag}} \sum_{k=1}^{t-1} x_{ij(t-k)}
\right).
\end{equation}

This structure maintains the ability to detect immediate effects while modeling lagged contributions with fewer parameters. These grouped specifications reduce dimensionality by imposing equality constraints within each block of lags, yielding interpretable lag-group effects while maintaining a flexible marginal mean structure.

In practice, grouping choices are often guided by subject‑matter expertise or exploratory patterns observed in a fully partitioned model. Such patterns can inform parsimonious semi‑partitioned structures without requiring identification of each individual lag‑specific coefficient. For example, clinicians may have strong prior knowledge about the temporal profile of a treatment or exposure—such as an expectation that a drug’s effect is primarily contemporaneous or persists only over the next one or two follow-up visits. In such settings, grouping lags into clinically meaningful blocks provides a principled way to reflect this knowledge in the marginal mean structure. When such information is unavailable, an exploratory fully partitioned model can be used to identify patterns in the lag-specific coefficients (e.g., clusters of similar magnitude or sign), which can then inform a more parsimonious semi-partitioned specification. This two-step strategy preserves the interpretability and stability of the grouped model while avoiding the estimation of many parameters and variance inflation inherent in the fully partitioned approach. In most applications, the primary parameter of interest is the contemporaneous effect $\beta^{|0|}$, whose interpretation remains robust even when the grouping structure for lagged effects is moderately misaligned with the underlying temporal pattern.

The specification of the grouping and associated lag structure is less critical in the marginal setting considered here than it would be in a conditional (e.g., autoregressive) model. The purpose of the semi-partitioned approach is to decompose the marginal covariate effect—typically aggregated over all lags in standard GMM \cite{lai2007marginal}—into interpretable components attributable to specific groups of lags, without requiring estimation of each individual lag-specific coefficient.

To ensure valid moment conditions, we define a matrix $M_j$ for each covariate $j$, where each element indicates the validity of a moment condition based on the classification framework of Lai and Small \cite{lai2007marginal}. This matrix is partitioned into $T$ components: $M_{j0}$ for immediate effects and $M_{jk}$ for lags $k=1,\dots,T-1$, each corresponding to a specific time difference $t-s=k$. Under a grouped‑lag specification, , for example Equation~\eqref{eq:groupedlag2}, the moment conditions associated with $M_{j1},\ldots,M_{j(T-1)}$ are combined according to the chosen grouping structure.

The estimating equations for continuous outcomes are constructed as:
\[
\frac{\partial \mu_{is}(\boldsymbol{\tilde\beta})}{\partial{\beta}_j^{[k]}} \left[ Y_{it} - \mu_{it}(\boldsymbol{\tilde\beta}) \right] 
= x_{ij(s-k)} \left[ Y_{it} - \mu_{it}(\boldsymbol{\tilde\beta}) \right],
\]
where $\boldsymbol{\tilde\beta}$ is the starting point for $\boldsymbol{\beta}$.

For binary outcomes, we use the logistic mean function 
\[
\mu_{it}(\boldsymbol{\beta}) = \frac{\exp(\eta_{it} (\boldsymbol{\beta}))}{1 + \exp(\eta_{it}(\boldsymbol{\beta}))},
\]
where
\[
\eta_{it}(\boldsymbol\beta)=\beta_0 + \sum_{j=1}^{J} \left( \beta_j^{|0|} x_{ij(t)} + \beta_{j,\text{lag}} \sum_{k=1}^{t-1} x_{ij(t-k)} \right)
\]
and the moment conditions become:
\[
\frac{\partial \mu_{is}(\boldsymbol{\tilde\beta})}{\partial \beta_{j}^{[k]} }\left[ Y_{it} - \mu_{it}(\boldsymbol{\tilde\beta}) \right] = x_{ij(s-k)} \mu_{is}(\boldsymbol{\tilde\beta}) \left[1 - \mu_{is}(\boldsymbol{\tilde\beta})\right] \left[Y_{it} - \mu_{it}(\boldsymbol{\tilde\beta})\right].
\]
These components are aggregated across all valid moment conditions and subjects to form the empirical moment vector $G(\bm{\beta})$. Using equation (1), the GMM estimator is then obtained by minimizing the standard quadratic form:

\begin{equation}
Q(\bm{\beta}) = G(\bm{\beta})^{'} \bm{W} G(\bm{\beta}).
\end{equation}

where $\bm{W}$ is a positive definite weighting matrix. Unlike earlier implementations that relied on derivative-free optimization (e.g., Nelder–Mead), we use the Broyden–Fletcher–Goldfarb–Shanno (BFGS) algorithm, which efficiently approximates the Hessian and improves convergence properties in nonlinear settings \cite{broyden1970convergence}.

\section{Simulations}
We conducted simulation studies to evaluate the finite-sample performance of the semi-partitioned GMM estimator in comparison to both the fully partitioned GMM and a benchmark GMM model using only lag-0 and lag-1 covariates. Our primary focus was on the coverage of 95\% confidence intervals and average interval lengths under different temporal correlation structures and covariate types. Across the first two settings, the grouped‑lag structure aligns with the underlying temporal pattern, whereas Setting III evaluates performance under moderate misalignment. This design allows us to assess robustness and to illustrate how grouped‑lag parameters summarize the combined influence of multiple lagged effects. These simulation settings are illustrative rather than exhaustive; the goal is to highlight stability--variance trade-offs across representative scenarios, not to claim dominance over fully partitioned models in all regimes. In these simulations, we use the grouped‑lag specification in Equation~\eqref{eq:groupedlag}.

\subsection{Setting I: Lag-Specific Effects with Stationary Covariates}
This scenario, adapted from Lai and Small \cite{lai2007marginal}, simulates a process where the outcome depends on both current and lagged values of a covariate. Specifically, we generate data from:
\begin{align}
&Y_{it} = \gamma_1 x_{it} + \gamma_2 x_{i,{t-1}} + b_i + e_{it}, \label{eq:AR1}\\
&x_{it} = \rho x_{i,t-1} + \epsilon_{it}, \notag
\end{align}
where $(b_i, e_{it}, \epsilon_{it})$ are mutually independent normal variables with mean zero and variances $4$,$1$ and $1$, respectively. The 
$x_{it}$ follow a (stationary) AR(1) process, i.e., $x_{i0} \sim N(0, \sigma_{\epsilon}^2/(1 - \rho^2))$. 

We consider time points $t=1,\ldots,T = 5$ for $i=1,\ldots,n=500$ individuals. The parameters $\gamma_1$, $\gamma_2$ and $\rho$ are 
set to $1$, $1$ and $.5$, respectively. Observe that in light of \eqref{eq:AR1} a partitioned GMM with the covariate at lags $0$ 
and $1$ represents the covariate-outcome relationship most accurately. In addition, we note that with $\mu_{it} = \gamma_1 x_{it} + \gamma_2 x_{i,{t-1}}$
\begin{equation*}
E[x_{is}(Y_{it} - \mu_{it})] = E[x_{is} (b_i + e_{it})] = 0, \quad \text{for any $1 \leq s \leq T$}, 
\end{equation*}
i.e., in this setting the covariate qualifies as Type I when both contemporaneous and first-lag effects are included in the model. However, we assume the lag structure is unknown to the analyst and fit both fully partitioned and semi-partitioned GMM models with the predictor $\mu_{it} = \sum_{s = 0}^{t-1} \gamma_{s+1} x_{i,t-s}$; in the semi-partitioned case, we employ the constraint $\gamma_3 = \ldots = \gamma_T$, which matches the true data‑generating process in this setting and corresponds to grouping all higher‑order lags into a single block. This grouping reflects a common applied scenario in which investigators expect the primary signal to lie in the contemporaneous and first‑lag effects, with remaining lags contributing a smaller, averaged influence.

Table \ref{sim:Setting1} presents the empirical coverage and average length of 95\% confidence intervals from 1,000 simulated datasets. For $\gamma_1$, all methods yield similar coverage just below the nominal level. For $\gamma_2$, the semi-partitioned GMM closely matches the oracle estimator, with improved efficiency over the fully partitioned approach. These results illustrate that when the true lag structure is simple, grouping higher-order lags yields performance comparable to the oracle model and can improve stability relative to the fully partitioned model.

\begin{table}[htbp]
\small
\begin{center}
\begin{tabular}{|c|c|c|c|c|}
\hline
               & \multicolumn{2}{c|}{$\gamma_1$}   & \multicolumn{2}{c|}{$\gamma_2$} \\ 
              \cline{2-3}  \cline{4-5} 
               & CI Coverage & Avg.~CI length &  CI Coverage & Avg.~CI length \\ 
Lag 1 GMM only &    .914      &  .0770         &    .94       &   .0844              \\ 
Semi-partitioned GMM &  .914    & .0820          &   .94        &  .0845               \\
(Fully) partitioned GMM  &  .916  & .0823          &  .944      &   .0872      \\
\hline
\end{tabular}
\end{center}
\caption{Coverage and average confidence interval (CI) length for Setting I (stationary covariate, known lag-1 structure)}\label{sim:Setting1}
\end{table}

\subsection{Setting II: Feedback with Type III Covariates}
The second simulation setting introduces a feedback loop between the outcome and the covariate process. Specifically, the data-generating model is as follows. 
\begin{align*}
&Y_{it} = \beta x_{it} + \kappa Y_{i,{t-1}} + u_{it}, \\
&x_{it} = \gamma Y_{i,t-1} + v_{it} 
\end{align*}
where the $u_{it}$ and $v_{it}$ are mutually independent normal random variables with mean zero and variances $\sigma_u^2$ and $\sigma_v^2$. 
This configuration induces a Type III covariate, where current covariate values are influenced by past outcomes. We set $n = 500$ and $T  = 5$ as before, and let $\beta = 1$, $\kappa = .4$, $\gamma = .2$, $\sigma_u^2 = \sigma_v^2 = 1$. As in Setting I, we consider predictors $\mu_{it} = \theta_1 x_{it} + \theta_2 x_{i(t-1)}$ (lag zero and one) and $\mu_{it} = \sum_{s = 0}^{t-1} \theta_{s+1} x_{i,t-s}$ with the constraint $\theta_3 = \ldots = \theta_T$ in the semi-partitioned case. Although a single‑lag model does not fully capture the feedback structure, higher‑order lag effects are negligible in the data‑generating process, making it a reasonable baseline for comparison.

Table \ref{sim:Setting2} summarizes the simulation results. All models attain coverage close to the nominal level. The semi-partitioned GMM shows substantial gains in efficiency over the single-lag GMM, and the fully partitioned model offers additional, though modest, gains at the cost of increased parameterization. These findings indicate that even under feedback, where valid moment conditions are limited, grouping higher‑order lags produces stable and interpretable estimates without compromising coverage.

\begin{table}[htbp]
\small
\begin{center}
\begin{tabular}{|c|c|c|c|c|}
\hline
               & \multicolumn{2}{c|}{$\theta_1$}   & \multicolumn{2}{c|}{$\theta_2$} \\ 
              \cline{2-3}  \cline{4-5} 
               & CI Coverage & Avg.~CI length &  CI Coverage & Avg.~CI length \\ 
Lag 1 GMM only &    .944      &  .177         &    .944      &   .768              \\ 
Semi-partitioned GMM &  .944    & .182          &   .944        &  .852               \\
(Fully) partitioned GMM  &  .948  & .188          &  .951      &   .923      \\
\hline
\end{tabular}
\end{center}
\caption{Coverage and average CI length for Setting II (feedback loop; Type III covariates)}\label{sim:Setting2}
\end{table}

\subsection{Setting III: Model Misspecification}
We revisit Setting I, modifying data generation per Eq.~\eqref{eq:AR1} as follows: 
\begin{align}
&Y_{it} = \gamma_1 x_{it} + \gamma_2 x_{i,{t-1}} + \gamma_3 x_{i,{t-2}} + \gamma_4 x_{i,{t-3}} +  \gamma_5 x_{i,{t-4}} + b_i + e_{it}, \label{eq:AR1_mod}\\
&x_{it} = \rho x_{i,t-1} + \epsilon_{it}, \notag
\end{align}
with $\gamma_3 = 0.3$, $\gamma_4 = 0.2$, and $\gamma_5 = 0.1$; all other quantities remain unchanged. This setting evaluates robustness when the true lag‑specific effects differ across lags but the semi‑partitioned model groups them. In such cases, the grouped parameter summarizes the combined influence of the unequal lag effects rather than recovering each coefficient individually, reflecting the parsimony of the grouped‑lag specification.

We consider the same approaches as before. Note that only the fully partitioned
GMM captures the modified lag structure, whereas the Lag 1 GMM and the semi-partitioned 
GMM that constrains the coefficients associated with lags beyond the first to be equal are 
incorrectly specified. Table \ref{sim:Setting3} summarizes the results for this modified
setting. The coverage rates for the semi-partitioned GMM undergo a slight
and moderate drop for the first and second coefficient, respectively. The semi‑partitioned estimator remains  substantially more robust than the single‑lag GMM, which exhibits severe undercoverage. The fully partitioned estimator performs best in this setting, as expected, but at the cost of estimating several additional parameters. When the grouping is misspecified, the semi‑partitioned estimator still provides stable summaries of the lag effects and offers a practical compromise between the oversimplified single‑lag model and the high‑dimensional fully partitioned model.

\begin{table}[htbp]
\small
\begin{center}
\begin{tabular}{|c|c|c|c|c|}
\hline
               & \multicolumn{2}{c|}{$\gamma_1$}   & \multicolumn{2}{c|}{$\gamma_2$} \\ 
              \cline{2-3}  \cline{4-5} 
               & CI Coverage & Avg.~CI length &  CI Coverage & Avg.~CI length \\ 
Lag 1 GMM only &    .342      &  .0834         &    .361      &   .0902              \\ 
Semi-partitioned GMM &  .906    & .0822          &   .820        &  .0848               \\
(Fully) partitioned GMM  &  .915  & .0822          &  .944      &   .0872      \\
\hline
\end{tabular}
\end{center}
\caption{Coverage and average CI length for Setting III (Model Misspecification).}\label{sim:Setting3}
\end{table}

\subsection{Summary}
Across the first two simulation settings, the semi-partitioned GMM model consistently maintains appropriate coverage and offers improved efficiency over the fully partitioned model. When the true lag structure is simple or when higher-order lags are weak, grouping yields estimates comparable to the oracle fully partitioned model but with slightly shorter confidence intervals (i.e., reduced sampling variability). When covariate effects vary over time, but most of the signal is concentrated in the immediate or first-lagged effect, the semi-partitioned approach provides a robust and parsimonious alternative. Even under misspecification, the semi‑partitioned estimator provides stable summaries of the lag effects and remains substantially more reliable than the single‑lag GMM, while requiring fewer parameters than the fully partitioned model. These results support its use in practical longitudinal settings where both interpretability and estimation stability are desired.

\section{Real Data Example}
We apply the semi-partitioned GMM framework to two real-world datasets to evaluate its performance relative to aggregated and fully partitioned GMM approaches. The datasets span both continuous and binary outcomes, feature multiple follow-up waves, and include time-dependent covariates, offering robust empirical validation of the proposed methodology. In both applications, our focus is on marginal, population-averaged effects rather than conditional dynamics, consistent with the modeling framework developed earlier.
 
\subsection{Osteoarthritis Initiative Data}

Our first application draws on the Osteoarthritis Initiative (OAI), a large, multi-center longitudinal study focused on knee osteoarthritis \cite{oai}. From this dataset, we selected 3,717 participants with complete measurements across three follow-up waves. The outcome of interest is the Western Ontario and McMaster Universities Osteoarthritis Index (WOMAC) disability score—a widely used continuous measure of pain, stiffness, and physical function. The score was averaged across both knees to provide a unified measure of lower-limb disability. Key predictors included sex (time-invariant) and two time-dependent covariates: age and body mass index (BMI), both of which are known risk factors for osteoarthritis progression. Age, BMI, and sex were included as predictors of WOMAC score in the model. Age and BMI were treated as time-dependent covariates, reflecting their potential to vary across follow-up periods and influence functional outcomes.

We applied all three GMM strategies—aggregated, fully partitioned, and semi-partitioned—to estimate marginal effects, and results are summarized in Table \ref{tab:linear}. Across all models, higher BMI and female sex were strongly associated with increased WOMAC scores, while age showed no significant contemporaneous effect. These results are consistent with prior literature and reinforce the validity of the marginal GMM approach in capturing established associations.

However, important differences emerged between models. The fully partitioned approach, while theoretically flexible, introduced several additional parameters for lagged age and BMI effects, none of which were statistically significant. This expansion led to increased standard errors and reduced estimation efficiency. These patterns are consistent with the challenges of estimating numerous lag-specific parameters in settings with modest follow-up, where higher-order lag effects are expected to be weak. In contrast, the semi-partitioned GMM model provided nearly identical point estimates for the immediate effects of BMI and sex, but achieved greater parsimony by grouping lagged effects. This grouping revealed that any delayed influence of age or BMI was negligible, while avoiding the increase in standard errors that arises when estimating many lag‑specific parameters.

These results illustrate that the semi-partitioned GMM model offers a more efficient and interpretable representation of longitudinal relationships in the presence of modest follow-up. It successfully retains the core advantages of full partitioning—namely, the ability to model lagged dynamics—while avoiding its pitfalls, particularly when lagged effects are weak or clinically unimportant. By decomposing the marginal effect into contemporaneous and grouped lag components, the semi-partitioned model captures immediate and delayed effects without overfitting. From a practical viewpoint, the grouped lag parameter should be interpreted as a stable summary of persistent exposure rather than a claim of equal lag effects across all lags.

These findings further corroborate prior evidence that links higher BMI and female sex with increased osteoarthritis-related pain and disability, while age—though often associated with worse outcomes—may exert a more modest or indirect influence when modeled alongside other covariates \cite{weiss2014relation, fang2015gender, ojeda2025sex}.

\begin{table}[htbp]
\centering
\resizebox{\textwidth}{!}{%
\begin{tabular}{|c|c|c|c||c|c|c|c||c|c|c|c|}
\hline
\multicolumn{4}{|c||}{No Partitioning} & \multicolumn{4}{c||}{Semi Partitioning} & \multicolumn{4}{c|}{Fully Partitioning} \\
\hline
 & Coef. & SE & p-val & & Coef. & SE & p-val & & Coef. & SE & p-val \\
\hline
(Intercept) & -13.8& 1.65& \textbf{<0.001} & (Intercept) & -14.44& 1.67& \textbf{<0.001} & (Intercept) & -14.46& 1.67& \textbf{<0.001} \\
Sex & 2.75& 0.33& \textbf{<0.001} & Sex & 2.60& 0.33& \textbf{<0.001} & Sex & 2.60& 0.33& \textbf{<0.001} \\
Age & -0.01& 0.02& 0.69& Age & 0.03& 0.02& 0.15& Age & 0.03& 0.02& 0.09\\
BMI & 0.68& 0.04& \textbf{<0.001} & BMI & 0.66& 0.04& \textbf{<0.001} & BMI & 0.65& 0.04& \textbf{<0.001} \\
- & - & - & - & Age (lags 1-2) & -0.01& 0.005*& 0.23& Age (lag 1) & -0.02& 0.01& 0.17\\
- & - & - & - & BMI (lags 1-2) & -0.01& 0.01& 0.61& Age (lag 2) & -0.002*& 0.01 & 0.86\\
- & - & - & - & - & - & - & - & BMI (lag 1) & -0.01& 0.02& 0.71\\
- & - & - & - & - & - & - & - & BMI (lag 2) & 0.002*& 0.02& 0.92\\
\hline
\end{tabular}%
}
\caption{GMM estimates for the WOMAC disability score across partitioning strategies}\label{tab:linear}
\begin{tablenotes}
\footnotesize
\item Note 1: Values are rounded to two decimal places, except for values $\leq$ 0.005 which are reported to three decimal places to preserve precision and are flagged with an asterisk (*).
\item Note 2: P-value of the significant variables are in bold
\end{tablenotes}
\end{table}

\subsection{Add Health Data}
The second application uses data from the National Longitudinal Study of Adolescent to Adult Health (Add Health), a nationally representative cohort tracking adolescents into adulthood over four waves of follow-up \cite{Addhealth}. The binary outcome is obesity status, defined according to standard BMI cutoffs. Time-dependent covariates include physical activity, number of hours of television viewing, depressive symptoms, and alcohol use. Race was included as a time-invariant covariate.

This dataset presents a more challenging modeling scenario due to the binary outcome, a higher number of follow-up waves, and the likelihood of temporal feedback and behavioral interdependence among covariates. In such settings, valid moment conditions are more limited—particularly for covariates influenced by prior outcomes—making model stability and parsimony especially important.

We applied all three GMM strategies and results are summarized in Table \ref{tab:logistic}. The fully partitioned GMM model, while flexible in principle, produced unstable and sometimes counterintuitive results. For example, the direction of association between TV hours and obesity reversed when partitioned across lags, and standard errors were inflated. These irregularities are consistent with the challenges of estimating many lag-specific parameters in a high-dimensional binary model, where multicollinearity and sparse information can distort inference. Together, these patterns suggest that the fully partitioned model overfits the data and was highly sensitive to correlations among lagged covariates. In contrast, the aggregated GMM model yielded stable estimates but collapsed all valid moment conditions into single coefficients, thereby obscuring important temporal nuance.

The semi-partitioned model outperformed both alternatives. It successfully recovered the expected relationships between increased screen time, reduced physical activity, and elevated obesity risk. Crucially, it also captured the persistent influence of depression across time, including statistically significant lagged effects—results that align with prior studies on adolescent mental health and obesity risk trajectories. By grouping lagged effects, the model stabilized estimation while preserving interpretable summaries of temporal influence. In this application, the grouped lag effect should be viewed as a coherent summary of longer-term exposure rather than an assertion that the individual lag-specific coefficients are identical.

These findings highlight the real-world value of the semi-partitioned GMM model. It handles the complexity of dynamic covariate behavior in longitudinal binary outcomes without overburdening the model with unnecessary parameters. The model's capacity to stabilize inference, recover lagged associations, and deliver clinically coherent results underscores its practical superiority in applied longitudinal research.

These results align with prior research demonstrating that depressive symptoms \cite{anderson2006depression, lindberg2020anxiety}, extended television viewing \cite{macdonell2023tv}, and reduced physical activity \cite{rauner2013physical, lundh2025activity} are associated with elevated obesity risk during adolescence and young adulthood. The semi-partitioned GMM model recovered significant immediate and lagged effects for all three covariates, with directionality consistent with the literature for depression and screen time. The grouped lag estimates provide a stable summary of persistent behavioral influence without requiring estimation of each individual lag coefficient. Notably, the model also detected a sign reversal for lagged physical activity, echoing prior findings that its long-term influence may be attenuated or confounded by behavioral and environmental factors \cite{lundh2025activity}.

\begin{table}[htbp]
\centering
\resizebox{\textwidth}{!}{%
\begin{tabular}{|c|c|c|c||c|c|c|c||c|c|c|c|}
\hline
\multicolumn{4}{|c||}{No Partitioning} & \multicolumn{4}{c||}{Semi Partitioning} & \multicolumn{4}{c|}{Fully Partitioning} \\
\hline
 & Coef. & SE & p-val & & Coef. & SE & p-val & & Coef. & SE & p-val \\
\hline
(Intercept) & -1.99& 0.13& \textbf{<0.001} & (Intercept) & -3.23& 0.15 & \textbf{<0.001} & (Intercept) & -2.59& 0.19& \textbf{<0.001} \\
Race & 0.07& 0.10& 0.46& Race & 0.18& 0.09& 0.05& Race & 0.11& 0.12 & 0.38 \\
TV hours & 0.01 & 0.002*& \textbf{<0.001} & TV hours & 0.01& 0.002*& \textbf{<0.001} & TV hours & -0.21& 0.10& \textbf{0.04}\\
Alcohol & 0.10& 0.06& 0.13& Alcohol & -0.05& 0.06& 0.45& Alcohol & 0.01& 0.004*& \textbf{<0.001} \\
Activity & -0.57 & 0.04& \textbf{<0.001} & Activity & -0.17& 0.03& \textbf{<0.001} & Activity & -0.41& 0.20& \textbf{0.04}\\
Depression & 0.77& 0.10& \textbf{<0.001} & Depression & 0.26& 0.10& \textbf{0.01}& Depression & -0.09& 0.05& 0.10\\
- & - & - & - & TV hours (lags1-3) & 0.006*& 0.001*& \textbf{<0.001} & TV hours (lag 1) & 0.004*& 0.01& 0.60\\
- & - & - & - & Alcohol (lags1-3) & 0.04& 0.05& 0.35& TV hours (lag 2) & 0.01& 0.01& 0.16\\
- & - & - & - & Activity (lags 1-3) & 0.13& 0.03& \textbf{<0.001} & TV hours (lag 3) & -0.003*& 0.01& 0.79\\
- & - & - & - & Depression (lags 1-3) & 0.40& 0.09& \textbf{<0.001} & Alcohol (lag 1) & 0.09& 0.20& 0.66\\
- & - & - & - & - & - & - & - & Alcohol (lag 2) & 0.04 & 0.21& 0.85\\
- & - & - & - & - & - & - & - & Alcohol (lag 3) & -0.32& 0.31& 0.29\\
- & - & - & - & - & - & - & - & Activity (lag 1) & -0.07& 0.10& 0.45\\
- & - & - & - & - & - & - & - & Activity (lag 2) & 0.14& 0.13& 0.26\\
- & - & - & - & - & - & - & - & Activity (lag 3) & -0.06& 0.18& 0.73\\
- & - & - & - & - & - & - & - & Depression (lag 1) & 0.46& 0.43& 0.28\\
- & - & - & - & - & - & - & - & Depression (lag 2) & 0.98& 0.52& 0.06\\
- & - & - & - & - & - & - & - & Depression (lag 3) & 1.03& 0.71& 0.15 \\
\hline
\end{tabular}%
}
\caption{GMM estimates for obesity status across partitioning strategies}\label{tab:logistic}
\begin{tablenotes}
\footnotesize
\item Note 1: Values are rounded to two decimal places, except for values $\leq$ 0.005 which are reported to three decimal places to preserve precision and are flagged with an asterisk (*).
\item Note 2: P-value of the significant variables are in bold
\end{tablenotes}
\end{table}

\subsection{Summary}
The applications presented in this section offer compelling evidence that the semi-partitioned GMM model is not only theoretically appealing but also practically advantageous across diverse longitudinal contexts. In the osteoarthritis study, which featured a continuous outcome and relatively few follow-up waves, the semi-partitioned model produced estimates that were statistically indistinguishable from those obtained with full partitioning, while avoiding the inflation of standard errors and overparameterization that accompanied the latter. The grouped lag structure provided a parsimonious yet informative decomposition of temporal influence.
Importantly, the semi-partitioned model retained the interpretability of lagged covariate effects, confirming that most of the signal was captured through contemporaneous associations—thus justifying the parsimony imposed by grouping lag terms.
In the Add Health study, which featured a binary outcome and a richer temporal structure, the benefits of the semi-partitioned approach became even more apparent. The fully partitioned model generated unstable and, at times, counterintuitive coefficient estimates, likely due to multicollinearity and overfitting introduced by estimating a large number of lag-specific parameters. In contrast, the semi-partitioned model not only stabilized inference but also correctly recovered known relationships between behavioral covariates (e.g., TV watching, depression, physical activity) and obesity. It achieved this while maintaining interpretability and avoiding the noise introduced by estimating numerous weak lag effects. By capturing both immediate and grouped lagged effects without requiring an unwieldy number of parameters, the model balanced flexibility with parsimony.

Across both examples, the semi-partitioned GMM demonstrated consistent, interpretable, and efficient performance. It captured the dynamic influence of time-dependent covariates without compromising statistical power or model stability. These empirical results reinforce the theoretical rationale for grouping: when higher-order lag effects are modest or diffuse, semi-partitioning provides a stable and scientifically meaningful summary of temporal dependence. These results underscore the practical advantages of the semi-partitioned framework in real-world longitudinal applications, particularly when the number of repeated measures is moderate and lag effects are present but not sharply distinct.
In sum, the empirical evidence validates the theoretical rationale for semi-partitioning: it preserves the essential temporal dynamics of the covariate–response relationship while guarding against the noise, inefficiency, and interpretive confusion that often accompany full lag disaggregation. The semi-partitioned GMM approach should therefore be viewed not as a compromise, but as a preferred default in many applied settings where modeling delayed covariate effects is important yet must be balanced with finite sample and design constraints.

\section{Discussion and Consequences}
Longitudinal data present complex analytical challenges when covariates evolve over time and exhibit feedback with outcomes. Accurately estimating marginal effects in such settings requires models that can flexibly account for both temporal variation and within-subject correlation. While fully partitioned GMM models provide the richest temporal resolution—estimating separate effects at each lag—they come with considerable trade-offs in terms of parameter proliferation, increased standard errors, and model instability. Conversely, aggregated GMM models collapse all valid moment conditions into a single estimate per covariate, gaining efficiency but at the cost of obscuring temporal dynamics entirely.
The semi-partitioned GMM model, as presented in this paper, offers a principled and practical solution that addresses the limitations of both extremes. By explicitly estimating contemporaneous effects and grouping lagged effects into structured sets, the model retains essential temporal information while avoiding estimation of several parameters. This balance is particularly valuable in applications with moderate follow-up, where higher-order lag effects are often weak but still scientifically relevant. Unlike fully partitioned models, which often suffer from inflated standard errors and convergence failures in the presence of moderate lag depth or limited sample size, the semi-partitioned model reduces the dimensionality of the parameter space. This improves statistical power and computational stability—without discarding the rich moment structure that distinguishes GMM from simpler alternatives such as GEE.

Moreover, while aggregated GMM models gain efficiency by design, they do so by averaging potentially heterogeneous effects. In doing so, they risk misleading conclusions in cases where immediate and delayed covariate effects operate in different directions or with varying strength. The semi-partitioned model mitigates this risk by preserving these distinctions—offering more accurate, interpretable estimates that better reflect the dynamic nature of the underlying processes.

Empirically, our simulation studies demonstrate that the semi-partitioned approach performs on par with or better than the fully partitioned model in terms of coverage and efficiency, particularly in scenarios with lag heterogeneity or feedback. The method also showed resilience when the grouping structure was misspecified, yielding stable estimates of averaged lag effects while avoiding the instability that can arise under full partitioning. In real data applications, the model recovered consistent, interpretable effects and avoided the instability and counterintuitive estimates that arose under full partitioning—particularly in the binary outcome setting. These findings reinforce the conclusion that semi-partitioning is not merely a compromise, but a superior modeling strategy for many practical settings.

From an applied perspective, the semi-partitioned model offers a pragmatic compromise for longitudinal studies with time-dependent covariates. It goes beyond a single aggregated effect that obscures temporal dynamics, yet avoids the instability and interpretational burden of estimating a separate coefficient for every lag. In our clinical examples, grouping lags into a small number of blocks yielded effect estimates that were both statistically stable and scientifically interpretable, allowing investigators to distinguish immediate from more persistent covariate effects without overfitting.

\subsection{Strengths}
The semi-partitioned GMM framework advances longitudinal marginal modeling by combining the interpretability of temporal effect estimation with the efficiency of parameter reduction. It offers a more stable and scalable alternative to full partitioning, while substantially improving the realism and informativeness of estimates compared to complete aggregation. Its ability to incorporate feedback structures and accommodate both continuous and binary outcomes further broadens its applicability.
Its adaptability across outcome types, incorporation of feedback effects, and reliance on stable optimization methods make it especially compelling for modern longitudinal health studies.

\subsection{Limitations}
Like all modeling frameworks, the semi-partitioned GMM has limitations. Its performance can still be affected by collinearity among covariates and their lags, particularly when many time-dependent predictors are included. Additionally, the model currently assumes complete data collected at regular intervals and does not incorporate corrections for missingness or uneven follow-up. Extensions to handle irregular observation times or informative dropout would further enhance its practical utility. The grouping of lagged effects, while beneficial for parsimony, may mask meaningful heterogeneity if one or two lag terms carry disproportionate influence.

\subsection{Future Directions}

While not critical to the approach proposed herein, we believe that further improvements can be achieved by devising principled strategies for selecting an optimal grouping of the $T-1$ lags under consideration. The proposed approach allows for an arbitrary fixed grouping of  into consecutive blocks (e.g., $\{1,2 \}$, \{3,4\}, \{5\} 
for $T = 6$). There are $1 + \sum_{k = 1}^{T-2} \binom{T-2}{k - 1}$ such groupings, which increases rapidly with $T$. For moderate $T$, it is still feasible to evaluate each of those groupings via a suitable model selection criterion; for larger values of $T$, penalized estimation is a possible approach.

In spirit, this search over candidate groupings parallels general-to-specific strategies in time-series econometrics (e.g., Hendry's general‑to‑specific (GETS) framework; see \cite{Hendry1995,Hendry2000}), although our focus remains on marginal mean structures rather than conditional dynamics. Related conditional models—such as MIDAS specifications—address similar questions about lag structure but target conditional forecasting relationships \cite{Ghysels2006,Andreou2010}. In contrast, the semi-partitioned GMM framework is tailored to longitudinal studies where the scientific interest lies in population-averaged covariate effects.

Developing principled, data-driven procedures for selecting or penalizing lag-group structures in the marginal GMM setting is the focus of ongoing work and will be presented in a separate paper.

Further methodological enhancements could include extensions to handle missing or irregular data, and tools for diagnostic checking and moment validity. These developments would extend the already substantial utility of the semi-partitioned model and reinforce its position as a preferred approach for longitudinal data with time-dependent covariates.

\section{Conclusions}
Modeling longitudinal data with time-dependent covariates presents persistent challenges, especially when covariate effects vary across time or involve feedback from prior outcomes. Existing GMM-based approaches have tended to polarize between two extremes: aggregated models, which gain efficiency by collapsing all valid moment conditions at the cost of obscuring temporal dynamics, and fully partitioned models, which offer maximal temporal specificity but often become overparameterized and unstable in real-world applications.
This paper introduces a semi-partitioned GMM framework that strategically balances flexibility, efficiency, and interpretability. By estimating immediate covariate effects separately and grouping lagged effects into structured blocks, our model preserves the ability to capture delayed covariate influence while mitigating the increase in standard errors and convergence issues that undermine fully partitioned approaches. The framework accommodates a range of covariate types, including feedback-driven structures (Type III), and supports both continuous and binary outcome modeling.
Through simulation studies, we demonstrated that the semi-partitioned model consistently achieves coverage and efficiency comparable to or exceeding fully partitioned models, with significantly fewer parameters. In two real-world applications—examining functional disability in osteoarthritis and obesity risk in adolescents—the semi-partitioned GMM produced stable, interpretable results, successfully recovering both contemporaneous and lagged associations while avoiding the instability and noise observed in the fully partitioned estimates.
Taken together, these findings support the semi-partitioned GMM not as a compromise, but as a preferred modeling strategy for many longitudinal applications. It offers a pragmatic solution for analysts seeking to balance temporal fidelity with statistical and computational tractability. By preserving key features of temporal dynamics without overfitting, this framework extends the practical scope of marginal GMM estimation and addresses a critical gap in the toolbox of longitudinal data analysis.

\newpage
\bibliographystyle{SageV}
\bibliography{yourbibfile}

\end{document}